\documentclass[letterpaper]{JHEP3}

\usepackage{epsfig}
\epsfclipon
\def\eq{\begin{equation}}
\def\eeq{\end{equation}}
\def\eqa{\begin{eqnarray}}
\def\eeqa{\end{eqnarray}}
\def\nn{\nonumber}
\def\ni{\noindent}

\title{The 6D SuperSwirl}

\author{S.~L.~Parameswaran$^{1,2}$ ,  G.~Tasinato$^{3}$  and I.~Zavala$^{4}$

\\
$^1$ Centre for Mathematical Sciences, DAMTP,
               University of Cambridge,\\
               Cambridge CB3 0WA UK.\\

$^2$ Perimeter Institute for Theoretical Physics,
                31 Caroline St. N., \\ Waterloo, ON, Canada, N2L 2Y5.

\\

$^{3}$  The Rudolf Peierls Centre for Theoretical  Physics,
 Oxford University,\\ 1 Keble Road, Oxford OX1 3NP, UK.

  \\

$^{4}$ IPPP, Physics Department, Durham University, \\
     Durham, DH1 3LE, UK.}

\abstract{We present a novel supersymmetric solution to a
nonlinear sigma
 model coupled to supergravity.  The solution represents a static,
supersymmetric, codimension-two object,  which is  different
 to the familiar cosmic strings.  In particular, we consider 6D chiral
 gauged supergravity, whose spectrum contains a number of hypermultiplets.
 The scalar components of the
hypermultiplet are charged under a gauge field, and supersymmetry
implies that they experience a simple paraboloid-like (or 2D
infinite well) potential, which is minimised when they vanish.
Unlike conventional vortices, the energy density of our
configuration is not localized to a string-like core. The
solutions have two timelike singularities in the internal
manifold, which provide the necessary boundary conditions to
ensure that the scalars do not lie at the minimum of their
potential. The 4D spacetime is flat, and the solution is a
continuous deformation of the so-called ``rugby ball'' solution,
which has been studied in the context of the cosmological constant
problem.  It represents an unexpected class of supersymmetric
solutions to the 6D theory, which have gravity, gauge fluxes and
hyperscalars all active in the background.}


\preprint{HEP-COLO-510, DAMTP-2005-79}

\begin{document}

\section{Introduction}

Sigma models in quantum field theory constitute one of the most
interesting theories with a wide range of applications in high
energy physics. One of the most remarkable examples of this is
represented in the linear sigma model theories, by the
(non)-abelian Higgs model, where a complex scalar field has a
Mexican hat potential. This theory has interesting static
solitonic solutions, which are the well known vortices (for a
review, see e.~g.~\cite{manton}).
  When coupled to gravity, these solitons give rise
to codimension-two objects, or cosmic strings, which have the
effect of producing a conical singularity in  spacetime, and which
could have been formed in early stages of the universe's
evolution, with important implications for cosmology
\cite{cosmic}. Recently, cosmic strings with a superstring origin
have also been considered due to its relevance for possible
connections between string theories and experiment \cite{Dcosmic}.
In this context, cosmic string solutions, or vortex solutions to
(non)-abelian Higgs models coupled to supergravity, have been
considered recently,  with interesting results \cite{dvali}. Other
global and local supersymmetric codimension-two  solutions to
abelian Higgs models have been considered in a different context
in three dimensions \cite{BBS,carlos}.  Moreover, string-like
solutions in nonlinear sigma models ({\it i.e.} those with
non-canonical kinetic terms) have long been known to exist
\cite{GOR}.

In this note, we would like to consider a particular nonlinear
sigma model, which is coupled to gauged 6D supergravity \cite{PS},
and study new examples of static, codimension-two, supersymmetric
configurations, different from cosmic strings.
 Nonlinear sigma models appear quite generically in this
context, because once scalar fields (which can arise in matter or
supergravity multiplets) are coupled to supergravity, they always
seem to form such structures \cite{diverse}.  This allows an
elegant geometrical treatment of what would otherwise seem a
highly intractable nonlinear system.
Moreover, gauged supergravities are coming to the fore in recent
years, since they describe the low energy effective theory of
string theory compactifications with fluxes.  There, the sigma
model describes the moduli of the compactification, and the fluxes
gauge certain isometries of the sigma model manifold, inducing a
scalar potential in the theory.  Indeed, in contrast to bosonic
sigma models, where there is no unique way to construct a gauge
invariant potential, for supersymmetric theories, supersymmetry (SUSY)
 is often powerful enough to determine the form of the potential uniquely.
In general, it will be different to the familiar Mexican hat shape.

In this paper, we concentrate our attention  on six
dimensional  chiral gauged supergravity \cite{RDSSS,RDSSSnew,MS,NS}, in
which a complex
  scalar field, $\phi$, has a paraboloid-like potential with a
minimum at $\phi=0$ \cite{DS}.
   We are interested in static configurations
which represent codimension-two objects in space-time and,
moreover, preserve some fraction of the supersymmetry of the
original system.

\smallskip
\smallskip

\ni {\it The model under consideration}

\smallskip

\ni
The 6D supergravity theory that we study here has received much
attention in the past,  mainly due to its interesting
phenomenological applications.
 For example, it shares many of the features of 10D supergravity
--- and so also of string vacua --- such as the existence of
chiral fermions \cite{RDSSS} with nontrivial Green-Schwarz anomaly
cancellation \cite{6dac}, as well as the possibility of having
chiral compactifications down to flat four dimensions \cite{RDSSS}.

 In its minimal form, the bosonic
spectrum contains the graviton, dilaton, and antisymmetric two-
and three-form field strengths.  The gauging  of a global
R-symmetry, together with supersymmetry, requires the presence of
a positive definite potential for the dilaton, with a Liouville
form.

The presence of anomalies can be avoided by adding to the spectrum
a number of hypermultiplets, suitably charged under the gauge
group \cite{RDSSS}, rendering the theory consistent also at the
quantum level.  The scalars of the hypermultiplets appear in the
potential, which has a minimum only when they vanish.

By switching on a magnetic
 monopole, the field equations admit a  background solution of the form $R^4
 \times S^2$, that preserves half of the supersymmetries of the vacuum, and
 stabilises one of the moduli in the spectrum.
    This configuration
was discovered by Salam and Sezgin (SS) \cite{ss}. It
 represents one of the simplest examples of  flux compactification down to four
 dimensions,  in which  partial moduli stabilisation  is achieved \cite{ABPQ1}.


Moreover, it has been  recently shown \cite{GGP}, that under
certain assumptions, an important property  of this theory is that
the SS flux compactification is {\it unique}. More precisely, the
authors of  \cite{GGP} showed that - for the minimal theory - if
one limits one's attention to vacua of the form $M_{4} \times
M_{2}$, where $M_{4}$ is a four dimensional space-time with
maximal symmetry, and $M_2$ a compact, regular two dimensional
manifold, then the theory admits a unique supersymmetric
configuration: the Salam-Sezgin one.

Thanks
 to the relative simplicity of the theory, it is interesting to
ask whether, by renouncing the assumptions of \cite{GGP}, other
potentially interesting supersymmetric  compactifications exist.
 For example, by considering different space-time factorisations,
the authors of
 \cite{GLPS} determined
 solutions of the form $AdS_3 \times S^3$,  as well as
dyonic string configurations;  generalisations of the latter have
been found in \cite{DS}.

Interestingly, new classes of supersymmetric solutions can
alternatively
  be found by renouncing the
 hypothesis of {\it
  regularity} of the internal manifold. The simplest example is the rugby ball
  vacuum, which is obtained by slicing a wedge from the sphere,
 and which allows in this way the presence
  of  conical
  singularities at the  poles. This vacuum provides a setting
for the supersymmetric large extra dimensions (SLED) brane world
scenario of \cite{cliff}. Here,  the conical singularities are
  interpreted as codimension-two brane worlds, where the Standard Model fields can be
  localized. The model has been introduced as a possible way to tackle the
  cosmological constant problem \cite{ABPQ} (for a non SUSY version,
see \cite{others}).
  More generally, this construction  shows how
 singular supersymmetric configurations can nevertheless be interesting
 as
 settings for brane world models, in which supersymmetry may help to ensure
 the stability
of the bulk  geometry, at least at the classical level.

\smallskip
\smallskip

\ni {\it Our results}

\smallskip

\ni
By renouncing the hypothesis of regularity, we show that further
supersymmetric vacua do exist,  preserving  four dimensional
maximal symmetry (they are 4D flat). They are obtained by  turning
on  the hyperscalars
 contained in the hypermultiplets, which are
 necessary for anomaly cancellation.  In this way, we have all
 kinds of fields in the theory active - gravity, gauge fields and
 scalars - which are consistent with the symmetries of the problem.
 The
 hyperscalar action corresponds to
 a nonlinear sigma model  defined on a non-compact, quaternionic
 manifold.  The scalar fields are coupled to gravity and to  gauge
 fields, and have a potential with a global minimum at zero~\footnote{This
result was first obtained in \cite{DS}.}.
  The study of
 sigma models in six dimensions, coupled only to gravity,
has been performed in \cite{ND,LP}, both in the supersymmetric and
 non-supersymmetric case.  However, as far as we are aware,
 our solutions constitute the first supersymmetric
 configurations in the full 6D gauged supergravity, in which the
 potential for the hyperscalar  fields, required
by supersymmetry, is included.
Given our field content, we are able to find the most general supersymmetric solution, which has maximal 4D symmetry and axial symmetry in the internal 2D space.

Interestingly,  the
 possibility to find a supersymmetric solution with the hyperscalars
turned on is naively {\it not} expected for this theory. Indeed,
the potential has a global minimum  at the origin. Nevertheless,
we show that a solution can be found, which preserves half of the
supersymmetries of the vacuum.
Similar to conventional vortices, the hyperscalars generate a
 configuration with a nonzero winding around the 2D manifold.  As usual,
this winding is induced by the coupling to the gauge field, as we
will discuss.  However, there are also significant differences to
the smooth
 vortex solutions that are generated by a Higgs potential and
spontaneous symmetry breaking.  In particular, vortices have a
well-defined core, at the center of which the scalar field sits
 at the top of its potential, where its amplitude is zero, $\phi=0$.
 Far away from the core, the scalar takes its minimum energy value.
 Especially for vortices generated by a local symmetry breaking, the
energy density of the vortex is localized near to the core.
 In contrast, in our solutions the energy density is not confined
 to a string-like region, and the scalar field is nowhere vanishing,
  which here means that
it does not  reach the minimum of its potential.
Moreover, the smooth central
core that arises in vortices, is replaced by two singularities,
which pinch off the internal manifold making its volume finite.

The singularities and winding are both needed for our
configuration, since otherwise the hyperscalars  would lie at the
minimum of their potential, where they vanish. The resulting
geometry is continuously connected to the rugby ball
configuration, when the hyperscalars are set to zero.
When the hyperscalars are switched on, the geometry deforms,  but
maintains nevertheless a ${\mathbb Z}_2$ reflection symmetry.
    The curvature singularities
at the poles transform  from  conical  to more serious ones, which
are the sources for the hyperscalar fields.
 Given the novelty of this codimension-two supersymmetric
configuration, we introduce a new name to define it; we call this
new object {\it the SuperSwirl}.

\smallskip
\smallskip
\ni {\it Outline}

\smallskip

\ni
The paper is organized as follows.
{\it Section} (\ref{secmodel}) contains a technical, but necessary
discussion of the explicit construction of the  action for the 6D
gauged supergravity. The hyperscalar part of the  action, in
particular, depends on the choice of the quaternionic manifold
that the hyperscalars parameterise. The reader interested only in
the final form of the sigma model that we consider, can jump this
section and
go directly to {\it Section} (\ref{seccondsusy}). 
In {\it Section} (\ref{seccondsusy}) we discuss in detail the 6D
nonlinear sigma model that we are interested in. We analyse the
conditions necessary to preserve some of the  supersymmetry,
taking a general ansatz for the fields involved. In {\it Section}
(\ref{secsusyconf}) we derive the supersymmetric solution, and
 discuss its properties.
In {\it Section} (\ref{secphysimpl}) we discuss the physical
implications of this configuration, and we conclude.


\section{The Six Dimensional Theory}\label{secmodel}

\subsection{Lagrangian, equations of motion, and susy transformations}

We consider the six-dimensional ${\mathcal N} = (1,0)$
 gauged supergravity constructed by
 Nishino-Sezgin (NS) in \cite{NS}.
The particle content of this theory consists of various fields: a gravity
multiplet $(e_M^{\it Q}, \, \Psi_{M_L}^A,\, B_{MN}^-)$; a tensor
multiplet $(B_{MN}^+,\,
 \chi_{_R}^A,\, \varphi)$;
Yang-Mills multiplets  $(A^{\hat I}_M,\, \lambda_{_L}^{{\hat I}
A})$;
 and the  hypermatter multiplets $(\Psi^a_{_R},\, \Phi^\alpha)$;
where $A=1,2$, $a=1,\dots,2n$, $\alpha=1,\dots, 4n$, with $n$ the
number of hypermultiplets. Also, $M=(\mu\,,m)$ are spacetime
indices in $d=6$ dimensions and ${\it Q}=(\hat\mu\,,q)$ are flat
tangent indices. The gravitino, dilatino and gaugini are all
$Sp(1)$ Majorana-Weyl spinors, and the hyperini are $Sp(n)$
Majorana-Weyl spinors.

The bosonic part of the corresponding Lagrangian is given by
\cite{NS}\footnote{See \cite{NS} for the fermionic part.}
\eqa \label{E:Baction}
    e^{-1} {\cal L}_B &=& \, \frac{1}{4 } \, R - \frac{1}{4 } \,
    \partial_{M} \varphi \, \partial^M\varphi  - \frac12 \, G_{\alpha \beta }(\Phi) \,
    D_M \Phi^\alpha  \, D^M \Phi^\beta  \cr
    && \qquad - \, \frac{1}{12}\, e^{2\varphi} \;
    G_{MNP}G^{MNP} - \, \frac{1}{4} \, e^{\varphi}
    \; F^{\hat I}_{MN}F_{\hat I}^{MN}
    -  \, \frac{1}{8} \, e^{-\varphi} \, v(\Phi) \, .
\eeqa

\ni
 Here as usual, $e = \sqrt{-det \,g}$, where $g_{MN}$ is the 6D
spacetime metric.  The Kalb-Ramond field strength is given by
$G_{MNP} =
    (\partial_M B_{NP} + F^{\hat I}_{MN} A^{\hat I}_P - \frac{1}{3}g'
    f^{ijk} A^i_M A^j_N A^k_P - \frac{1}{3} g f^{IJK} A^I_M A^J_N
    A^K_P) + {\rm 2 \,\,\,perms}$.  The index $\hat I $ runs over the adjoint of
    $Sp(n) \times Sp(1)$, and so can be subdivided into the adjoint of
    $Sp(n)$: $\hat I = I = 1, \dots, n(2n+1)$, and the adjoint of
    $Sp(1)$: $\hat I = i = {\it 1, 2, 3}$.  The structure constants of the
    gauge group are then labelled by $f^{IJK}$ and $f^{ijk}$.
Supersymmetry requires
 that the hyperscalars parameterize a quaternionic manifold:
\eq \frac{G}{H} = \frac{Sp(n,1)}{Sp(n) \times Sp(1)} \, , \eeq
whose metric is
    $G_{\alpha \beta }(\Phi)$.  Thus the index $\alpha = 1, \dots 4n$ can
be  interpreted as   the curved index on this target space
manifold.

The geometry of the target manifold can be described by the
Maurer-Cartan form, which is constructed from the
coset-representative, $L$: \eq L^{-1} \partial_{\alpha} L =
{\mathcal W}^{\hat X}_{\alpha} T^{\hat X} + V^{aA}_{\alpha} T_{aA}
\, . \eeq Here, $T^{\hat X}$ ($\hat X = x\,,X$) and $T_{aA}$ are
the anti-hermitian generators of $Sp(n) \times Sp(1)$ and the
coset, respectively. Then, ${\mathcal W}^{\hat X}_{\alpha}$
transforms as the spin-connection on the target manifold $G/H$,
and $V^{aA}_{\alpha}$ is the vielbein, carrying the tangent space
indices $a=1, \dots, 2n$ and $A = 1,2$, which run over the
fundamental of $Sp(n) \subset H$ and $Sp(1) \subset H$
respectively.
 The scalar potential then takes the form

\eq v = C^{x \hat I} C^{x \hat I} \eeq with \eq\label{E:defC} C^{x
\hat I}  =
      \left\{ \begin{array}{l}
          C^{xI} = g {\mathcal W}^x_{\alpha} \xi^{\alpha I} \\
          C^{xi} = g' ({\mathcal W}^x_{\alpha} \xi^{\alpha i} -
          \delta^{xi})\end{array} \right. \, .
\eeq

This prepotential $C^{x{\it \hat I}}$ can also be calculated
directly from the coset representative \cite{PS,PS1}
as follows: \eq C^{x \hat I} = 2\,g' \,\left( L^{-1} {T}^{\hat I}
L \right)_{AB} T^x_{AB}\,. \eeq

\ni  The   Killing vectors of the scalar manifold are $\xi^{\alpha
\hat
    I}(\Phi) = T^{\hat I \alpha}_{\beta}\Phi^{\beta}$ (the
$T^{\hat I \alpha}_{\beta}$ are anti-hermitian generators for the
groups $Sp(n)$ and $Sp(1)$ in $G$).
     Finally, the covariant derivative of the scalars is given by:
\eq D_M \Phi^{\alpha} = \partial_M \Phi^{\alpha} - g'A^i_{M}
\xi^{\alpha i}
              - g A^I_{M} \xi^{\alpha I}
\eeq

\ni Here, $g'$ is the gauge coupling of the $Sp(1)$ group, and $g$
that of the $Sp(n)$ group. The explicit definitions
    in terms of the scalars depends on the choice of coset
    representative, to be discussed in the next subsection.

The bosonic equations of motion  derived from the corresponding
action are:
\eqa \label{E:Beom}
    &&\Box \, \varphi - \frac{1}{3} \, e^{2 \varphi}
    \, G_{MNP} \, G^{MNP} - \frac12 \, e^{\varphi} \; F^{\hat I}_{MN}
    F^{MN}_{\hat I} + \frac{1}{4} \, e^{-\varphi} \, v(\Phi) = 0 \nn\\
    &&D_M \Bigl( e^{2\varphi} \, G^{MNP} \Bigr) = 0  \\
    &&D_M \Bigl( e^{\varphi} \, F^{MN}_{\hat I} \Bigr) - e^{2\varphi} \,
    G^{MPN} \, F_{{\hat I} M P} + \hat g
\, g^{MN}(D_M \Phi^\alpha) \,\xi_{\alpha
    \hat I}= 0 \nn \\
    && D^M D_M \Phi^\alpha  + \Gamma^{\alpha}_{\beta\gamma} D^M
    \Phi^{\beta} D_M \Phi^{\gamma} - \frac{1}{8} G^{\alpha \beta }(\Phi) \, v_{\beta }(\Phi) \, e^{-\varphi} = 0 \nn\\
    &&R_{MN} - \partial_M\varphi \, \partial_N \varphi - 2
    G_{\alpha \beta }(\Phi) \, D_M \Phi^\alpha  \, D_N \Phi^\beta  -
    e^{2\varphi} \, G_{MPQ} \, {G_N}^{PQ} \nn\\
    && \qquad \qquad \qquad - 2\, e^{\varphi} \, F_{MP}^{\hat I}
    {F_{{\hat I} N}}^P + \frac{1}{2}\,  (\Box \varphi )\, g_{MN} = 0 , \nn
\eeqa
where $v_\beta  = \frac{\partial v}{\partial \Phi^\beta }$.  Also,
$\hat g = g \, {\rm or} \, g'$ depending on the gauge sector.
 The supersymmetry transformation rules for the fermions are (up to
fermion bi-linears):
\eqa && \delta\chi^A=
-\frac{1}{2}\partial_M\varphi\Gamma^M\epsilon^A +
  \frac{1}{12}e^{\varphi} G_{MNL}\Gamma^{MNL}\epsilon^A \label{susy1}\\
&& \delta\Psi^A_M = D_M\epsilon^A + \frac{1}{24}e^{\varphi}
  G_{NLR}\Gamma^{NLR}\Gamma_M\epsilon^A \label{susy2}\\
&& \delta \lambda^{A\hat I} =  \frac{1}{2\sqrt{2}}e^{\varphi/2}
 F_{MN}^{\hat I}\Gamma^{MN} \epsilon^A -\frac{1}{\sqrt{2}}e^{-\varphi/2}
        C^{x\hat I}T^{xA}_B\epsilon^B \label{susy3} \\
&& \delta\Psi^a =
(D_M\Phi^\alpha)\,V^{aA}_{\alpha}\Gamma^M\epsilon_A \,.
\label{susy4} \eeqa
Recall that all spinors are symplectic-Majorana Weyl.
  The
gravitini, Killing spinor, gaugini, and dilatini are all in the
fundamental of $Sp(1) \subset H$, whereas the hyperini are in the
fundamental of $Sp(n) \subset H$.  The $T^{xA}_B$ are the
generators of $Sp(1) \subset H$.  The gaugini are also in the
adjoint of $Sp(n) \times Sp(1) \subset G$.

The covariant derivative acting on  the Killing spinor is given
by: \eq D_M \epsilon^A = \partial_M \epsilon^A + \frac{1}{4}
\omega_M^{\it P Q} \Gamma_{\it P Q} \epsilon^A + g' A_M^{i}
\delta^{ i x} T^{xA}_B \epsilon^B + (D_M \Phi^{\alpha}) {\mathcal
W}_{\alpha}^x T^{xA}_B \epsilon^B \eeq

\ni where, as before, ${\it P,\,Q}$ are flat tangent indices on
the spacetime manifold. Note also that the last term, involving a
coupling with the scalars, is due to the fact that the Killing
spinor behaves as a section on the $Sp(1)$-bundle of the target
manifold.

\bigskip

\subsection{Parameterization of the Target Manifold}

We can express the scalars $\Phi^{\alpha}$ ($\alpha = 1, \dots,
4n$), as an $n$-component quaternionic vector, $t^p$ ($p = 1,
\dots, n$).  So, for example, $t^{p=1} = \Phi_1 1 + \Phi_2
\hat{\i} + \Phi_3 j + \Phi_4 k$, where we use the following  $2
\times 2$ basis of quaternions:

\eqa \label{E:qbasis} \hat{\i} &=&  \left(
     \begin{array}{cc}
      -i & \,\,0 \\
      \,\,0 & \,\,i  \end{array}
     \right) \,= \,-i \sigma_{3} \qquad \qquad
j = \left(
     \begin{array}{cc}
      0 & -1 \\
      1 & \,\,0 \end{array}
     \right)\,=\,-i \sigma_{2} \cr \nn \\
k &=& \left(
     \begin{array}{cc}
      \,\,0 & -i \\
      -i & \,\, 0 \end{array}
     \right) \,=\,-i \sigma_{1} \qquad \qquad 
1 = \left(
     \begin{array}{cc}
      \,\, 1 & \,\,0 \\
      \,\, 0 & \,\,1 \end{array}
     \right)
\eeqa

We need to choose a specific parameterization of the target
manifold, in order to have explicit expressions for the metric,
$G_{\alpha \beta }(\Phi)$, and potential, $v(\Phi)$, which appear
in the field equations. A choice of coset representative, $L$ --
where $L$ is an $Sp(n,1)$ valued matrix --
 is sufficient to define all
necessary quantities. Following \cite{NS},   we choose this matrix
to be: \eq L = \gamma^{-1} \left(
     \begin{array}{cc}
      1 & t^{\dagger} \\
      t & \Lambda(t) \end{array}
     \right)
\eeq where \eq \gamma = (1-t^{\dagger}t)^{1/2} \, ,
\phantom{000000} \Lambda(t) =
      \gamma(I-tt^{\dagger})^{-1/2}
\eeq Here, $I$ is the $n\times n$ unit matrix, and $^{\dagger}$
refers to
      matrix transposition and quaternionic conjugation ($a + b\hat{\i} + cj
      + dk \rightarrow a - b\hat{\i} - cj -dk$).
The Maurer-Cartan form $
      (L^{-1}\partial_{\alpha}L)$ can now
      be decomposed as:
\eq L^{-1} \partial_{\alpha} L = \left(
     \begin{array}{cc}
      {\mathcal W}_{\alpha}^{AB} & V_{\alpha}^{\dagger Ab} \\
      V_{\alpha}^{aB} & {\mathcal W}_{\alpha}^{ab} \end{array}
     \right)
\eeq where ${\mathcal W}_{\alpha}^{AB} = {\mathcal W}_{\alpha}^x
T^{x \, AB}$
 and ${\mathcal W}_{\alpha}^{ab} = {\mathcal W}_{\alpha}^X T^{X \, ab}$ are
the $Sp(1)$ and $Sp(n)$ connections, and $V_{\alpha}^{aB}$ is the
pullback of the vielbein.
From these expressions it follows that: \eqa \label{E:targetgeom}
{\mathcal W}_{\alpha}^{AB} &=& \frac{1}{2} \gamma^{-2} \left(
\partial_{\alpha}
      t^{\dagger} \, t - t^{\dagger} \partial_{\alpha}t \right) \\ \nn
{\mathcal W}_{\alpha}^{ab} &=& \gamma^{-2} \left(
-t\partial_{\alpha}
      t^{\dagger} + \Lambda \partial_{\alpha}\Lambda + \frac12
      \partial_{\alpha}(t^{\dagger}t) I \right) \\ \nn
V_{\alpha}^{aA} &=& \gamma^{-1} \left( I - tt^{\dagger}
\right)^{-1/2}
      \partial_{\alpha}t
\eeqa

Choosing to turn on only two ``real'' components of the full
quaternion
 ($\Phi_1$, $\Phi_3$),
and using the above relations, the relevant
 metric components can be calculated from:
\eq G_{\alpha\beta} = V_{\alpha}^{aA} \epsilon_{ab} \epsilon_{AB}
V_{\beta}^{bB} \eeq and are given by:
\eqa \label{E:simplemetric11}
G_{11} &=& V_{1}^{aA} V_{1 aA} = \frac{2}{(1-\Phi_1^2 - \Phi_3^2)^2} \cr \nn\\
G_{33} &=& V_3^{aA} V_{3 aA} = \frac{2}{(1-\Phi_1^2 - \Phi_3^2)^2} \cr \nn\\
G_{13} &=& V_1^{aA} V_{3 aA} =0 \eeqa

\ni Here, we have used that the flat indices are raised and
lowered with the metric $\epsilon_{AB} =\epsilon^{AB} =
[(0,1),(-1,0)]$ and $\epsilon_{ab}$.  Also, we split the indices
$a = 1,\dots, 2n$ into $a = pA'$, with $p = 1 \dots n$ and $A' =
1,2$, and use $\epsilon_{ab} = {\sf 1}_{pq} \otimes
\epsilon_{A'B'}$.

In order to calculate the potential, we use an explicit form for
 the Killing vectors.  For simplicity, we will take the $Sp(n)$ gauge
coupling to zero in what follows, that is, $g=0$.
Moreover, we  consider only the gauging of the
$U(1)$ subgroup, that is, from now on we take $i={\it 1}$.
The Killing vectors are then:
\eqa
\xi^{{\it 1} \alpha} &=& T^{{\it 1} \alpha}_{\beta}\Phi^{\beta} \nn \\
  &=& T^{{\it 1} \alpha}_1 \Phi_1 + T^{{\it 1} \alpha}_3 \Phi_3
\eeqa
We choose the conventions for the relevant generator,
$T^{{\it 1} \, \alpha}_{\beta}$: $T^{{\it 1} \alpha}_{\beta} =
T^{{\it 1} A}_{B} \otimes {\sf 1}^a_b$, with $T^{{\it 1} A}_B =
1/2\,([0,-1],[1, 0])$, $A = 1,3$. Now, we can compute the $C$
function from the definition in
 (\ref{E:defC}), using the explicit values for ${\mathcal
 W}_\alpha^x$,
obtained from (\ref{E:targetgeom}) in terms of the two non zero
fields, $(\Phi_1,\,\Phi_3)$. We find:
\eqa && C^{x{\it 1}} =-\frac{g'}{1-\Phi_1^2 - \Phi_3^2}
\qquad\qquad
    {\rm for} \qquad x={\it 1}, \label{E:explicitC} \nn \\ \nn \\
&& \hskip0.8cm = 0 \hskip4cm {\rm otherwise } \eeqa

\ni The potential is then
\eq v = C^{x {\it 1}} C^{x{\it 1}} =  \frac{g'^2}{(1-\Phi_1^2 -
\Phi_3^2)^2}\,.
\eeq\\
We have found in this way all the quantities that characterize our
nonlinear sigma model. %
Using the above relations, we can now write explicit expressions
for the action, equations of motion and supersymmetry
transformations in terms of a
single complex scalar field defined as
\eqa\label{complexPhisdef} \phi = \Phi_1 + i\Phi_3\,,
 \qquad \qquad \phi^\star= \Phi_1 - i\Phi_3\,,
\eeqa We do this in the next section.

\section{The Model}\label{seccondsusy}

 The bosonic action for our 6D nonlinear sigma model
(\ref{E:Baction}), in terms of (\ref{complexPhisdef}), reduces to
%
 \eqa \label{E:Baction1}
     e^{-1} {\cal L}_B = \, \frac{1}{4 } \, R - \frac{1}{4 } \,
     \partial_{M} \varphi \, \partial^M\varphi
 - \frac{1}{(1-|\phi|^2)^2} \,
     D_M \phi \, D^M \phi^\star  
- \, \frac{1}{4} \, e^{\varphi}
     \; F_{MN}F^{MN}
    -\, \frac{1}{8} \,\frac{g'^2\, e^{-\varphi}}{(1-|\phi|^2)^2} \, \nn
\, ,  \\
 \eeqa
The equations of motion for our system then become:
\eqa\label{E:expliciteoms} &&\Box \, \varphi =\frac12 \,
e^{\varphi} \; F_{MN}
    F^{MN} - \frac{g'^2}{4} \,
    \frac{e^{-\varphi }}{(1-|\phi|^2)^2} \,,\nn\\
&& D_M \Bigl( e^{\varphi} \, F^{MN} \Bigr) =\frac{i
g'}{2(1-|\phi|^2)^2}\,
g^{MN}(\phi^\star D_M\phi - \phi D_M\phi^\star) \,,\nn \\
&& R_{MN} = \partial_M\varphi \, \partial_N \varphi + \frac{2}{(1-
|\phi|^2)^2} (D_M \phi D_N \phi^\star +D_M
 \phi^\star D_N \phi) \nn\\
        && \hskip5cm  + \,2 e^{\varphi} \, F_{MP}
    F_N^{\,\,\,P} - \frac12 \,  (\Box \varphi )\, g_{MN}\,,   \nn \\
&& D^M D_M \phi  + \frac{2\phi^\star}{(1-|\phi|^2)}
    D^M \phi D_M \phi =
    \,\frac{g'^2 e^{-\varphi}}{4} \frac{\phi}{(1-|\phi|^2)}\,.
\eeqa

\bigskip

\ni Here we also have to add an equation for the complex conjugate
of the scalar field.
 In terms of the complex field, the scalar manifold metric is:

 \eq d\sigma^2 = 2\frac{d\phi d\phi^\star}{(1-|\phi|^2)^2}\,. \eeq

\ni
The covariant  derivatives are given by\footnote{We computed
the covariant derivatives for $\phi$ using first the definition of
$D_M\Phi^\alpha$ in the previous section  and changing to
$\phi,\,\,\phi^\star$ notation.}:
\eqa\label{DMdef1}
&& D_M \phi = \partial_M \phi - \frac{ig'}{2}
A_M \phi\,, \\ \label{DMdef2} && D^M D_M \phi = \nabla^M D_M \phi
- \frac{ig'}{2} A^M\,D_M\phi \,,
\eeqa

where $\nabla_M$ is the covariant derivative with respect to the
metric:

\eq \nabla^M D_M \phi = \frac{1}{\sqrt{g}}
        \partial_M(\sqrt{g}\,g^{MN}\,D_N\,\phi)\,,\nn
\eeq
and equivalently for the complex conjugate field
$\phi^\star$.

\smallskip

The supersymmetry transformations (\ref{susy1}-\ref{susy4})
can be written as:
\eqa
&& \delta\chi= -\frac{1}{2}\partial_M\varphi\Gamma^M\,\epsilon\label{susyI}\\
&& \delta\Psi_M = D_M\epsilon \label{susyII}\\
&& \delta \lambda =  \frac{e^{\varphi/2}}{2\sqrt{2}}
 F_{MN}\Gamma^{MN} \,\epsilon +  \frac{i}{\sqrt{2}}
\frac{g'\,e^{-\varphi/2}}{(1-|\phi|^2)} \,\epsilon \label{susyIII} \\
&& \delta\Psi=
\frac{1}{2\,(1-|\phi|^2)}D_M\phi\,\Gamma^M\,\epsilon \,.
\label{susyIV} \eeqa
Here, all spinors are complex-Weyl and we have defined  them as
$\epsilon= \epsilon_1 + i\epsilon_2 $.

\subsection{Supersymmetry conditions }

We consider the most general ansatz consistent with 4D maximal
symmetry. Thus, we take:
\eqa\label{E:ansatz3}
ds^2= e^{2W(z,\bar z)}\, g_{\mu\nu} \, dx^\mu dx^\nu +
e^{2B(z,\bar z)}dzd\bar z\,,\nn
\eeqa
\eq
F^{\hat I}_{MN} = F_{mn}(z,\bar
z)\,,  \qquad \varphi=\varphi(z,\bar z)\,, \qquad  \phi =
\phi(z,\bar z)\,.
\eeq

\noindent where $g_{\mu\nu}$ is the 4D metric on de Sitter,
Minkowski or anti-de Sitter spacetime, and $z,\,\bar z$ are
complex coordinates in the internal 2 dimensions.  All other
fields are zero.

We now look at the supersymmetry transformations, to find what
conditions must be satisfied by the fields in order to ensure that
the system preserves some fraction of the total supersymmetry.
Since all the fermion fields vanish, we need only concern
ourselves with the transformation laws of the fermions.

\smallskip

\ni {\underline{$\delta\chi=0$}}: Plugging our ansatz above  into
the SUSY transformations, we see immediately from  the dilatino
condition that the dilaton must be constant: \eq
 \varphi=\varphi_0\,= \,{\rm constant}\,.
\eeq

\bigskip

\ni {\underline{$\delta\lambda=0$}}: \noindent From the  gaugino
condition, we have:
\eq \frac{1}{2}F_{MN}\Gamma^{MN}\,\epsilon=
        -i \frac{g'\, e^{-\varphi_0}}{(1-|\phi|^2)} \, \epsilon\,.
\eeq

\ni Writing $F_{q\bar q} =
     if(z,\bar z)\varepsilon_{q\bar q}$, with $f=f^\star$, and where $q,\,\bar q$
are the internal, flat, complexified indices and
$\varepsilon_{q\bar q} = \varepsilon^{q\bar q} = 1$, gives us:
\eq   
f(z,\bar z)\Gamma_{\bar qq}\,\epsilon =
      -\frac{g'\, e^{-\varphi_0}}{2}\frac{1}{(1-|\phi|^2)}\,\epsilon \,.
\eeq
In order to satisfy this condition, we   impose the following
projections on the spinors
\eq\label{projectionII}
\Gamma_{\bar qq} \,\epsilon = \epsilon \,,
\qquad \qquad  \Gamma_{\bar qq} \,\epsilon^\star = -\epsilon^\star
\eeq
which imply the following condition between the flux and the
potential
\eq\label{varconstrI} 
f=
-\frac{g'}{2}\frac{e^{-\varphi_0}}{(1-|\phi|^2)}  
\eeq

\smallskip

\ni The projection condition (\ref{projectionII}), breaks one
half of the 6D supersymmetries, thus leaving ${\mathcal N}=1$
from a four dimensional point of view.

\bigskip

\ni {\underline{$\delta\Psi=0$}}: \noindent The SUSY condition for
the hyperino, gives us immediately
\eq D_{\bar z} \,\phi \,\epsilon =0 \,, \qquad \qquad D_{z}
\,\phi^\star \,\epsilon^\star =0\,, \eeq
where $D_{\bar z} \phi = \partial_{\bar{z}} \phi - i\,\frac{g'}{2}
A_{\bar{z}} \phi$. Because we require non-vanishing spinors, this
implies that the complex scalar field must be {\it covariantly
holomorphic}, that is

\eq\label{constraint1}
D_{\bar z} \,\phi\,=\,0\,, \qquad
\qquad  D_{z} \,\phi^\star\, =\,0\,
\eeq

\bigskip

\ni {\underline{$\delta\psi_M=0$}}: \noindent The last SUSY
transformation is that for the gravitino. In order to compute
this, we need the values of the space-time spin connection. For
example, when the non-compact directions are Minkowski
$g_{\mu\nu}=\eta_{\mu\nu}$, the nonzero components are given by:
\eqa \label{spinconnection}
&& \omega^{\hat\mu q}_{\,\,\,\,\,\,\mu} =
\sqrt{2}\,e^{W-B}W_{\bar z}\,\delta^{\hat\mu}_{\,\,\mu}\,,  \qquad
\omega^{\hat\mu \bar q}_{\,\,\,\,\,\,\mu} =
    \sqrt{2}\,e^{W-B}W_{z}\,\delta^{\hat\mu}_{\,\,\mu}\,, \nn\\
&& \omega^{\bar qq}_{\,\,\,\,\,\,z} = -B_z \,,\qquad \qquad \qquad
\,\,\,\,
    \omega^{\bar qq}_{\,\,\,\,\,\,\bar z} = B_{\bar z}\, ,
\eeqa
where $\hat\mu$ and $q(\bar q)$ are flat indices. Assuming that
the spinor $\epsilon$ is a function only of $z,\,\bar z$, from the
$M=\mu$ component of the gravitino equation; \eq
\label{E:deltapsimu} \omega^{\,\,\,{\it PQ}}_\mu\,\Gamma_{\it
PQ}\,\epsilon=0\,, \eeq we can see that, for a 4D Minkowski
solution, $$W(z,\bar z) \,=\,{\rm constant}\,.$$

\smallskip
\noindent
On the other hand, for de Sitter or anti-de Sitter 4D spacetimes,
the condition (\ref{E:deltapsimu}) imposes additional projection
conditions on the Killing spinor, which break the remaining ${\cal
N}=1$ supersymmetry in 4D.  For example, for the AdS metric in
Poincar\'e coordinates: $ds^2 = e^{2W} \left(l^2/v^2 \right)
\left( -dt^2 + dv^2 + dx^2 + dy^2 \right) + e^{2B} dz d\bar z$, we
again arrive at $W(z,\bar z) \,=\,{\rm constant}$, but furthermore
find that we must impose: \eq \Gamma_{\hat \mu \hat v} \epsilon= 0
\eeq These projections break a further half of the original
supersymmetries.
  In order to avoid this situation, which would leave us with less than one
 supersymmetry at the four dimensional level, we are
forced to consider a flat 4D spacetime.

\smallskip
Now considering the $M=m= z,\bar z$ components of the gravitino
transformation, we find:
\eqa \label{E:deltapsim}
 \partial_m\epsilon + \frac12 \omega_{m}^{\,\,\bar{q} q}\Gamma_{\bar{q} q}
\,\epsilon +\frac{i g' A_{m}}{2}\,\epsilon +
 \frac{1}{2\,\left(1-|\phi|^{2}\right)}\Big[ \phi D_m\phi^{\star}
-\phi^{\star} D_m\phi \Big]\,\epsilon =0\,,
\eeqa

\ni and its complex conjugate. One can see  from these equations
that
\eq\label{spinors} \epsilon \,\epsilon^\star = constant \,. \eeq

\smallskip

Equation (\ref{E:deltapsim}) for $\epsilon$ is the last equation
that must be satisfied
 in order to preserve supersymmetry.  To ensure that such a Killing
spinor exists locally,
  it is sufficient to impose  the following
integrability  condition:

\eq\label{integra} 0 \,=\, \left[ D_{z}, D_{\bar{z}} \right]
\,\epsilon^A \,=\, \left( B_{z\bar z}
    +\frac{D_{z}\phi D_{\bar{z}}\phi^{\star}}{\left(1-|\phi|^{2}\right)^2}
+\frac12 e^{2B+\varphi_0}\,f^2 \right) \epsilon^A \,, \eeq where
in the last equality we have applied the conditions that emerge
from the preceding transformations. We must now check whether the
above constraints are consistent with the equations of motion.

\subsection{The equations of motion}

The supersymmetry constraints allow us to obtain and satisfy the
equations of motion. Indeed, the equation  for the
dilaton, $\varphi$, requires:
\eq\label{dilaeq1} \frac12 F_{MN}F^{MN} = \frac{g'^2}{4}
    \frac{e^{-2\varphi_0}}{(1-|\phi|^2)^2}\,.
\eeq
Plugging the value of $F_{mn}$ into the equation above gives us

\eq\label{fincont}
 f^2 = \frac{g'^2}{4}
     \frac{e^{-2\varphi_0}}{(1-|\phi|^2)^2}\,,
\eeq
which is precisely (\ref{varconstrI}).
The equation of motion for $\varphi$ is consequently
satisfied.

\smallskip

The equation of motion for the gauge field is:
\eq \frac{1}{\sqrt{g}}\partial_M(\sqrt{g}\, e^{\varphi_0} F^{MN})
= \frac{i g'}{2(1-|\phi|^2)^2}\, g^{MN}(\phi^\star D_M\phi - \phi
D_M\phi^\star) \,. \eeq

\ni Taking into account that $F^{z\bar z}=- 2if e^{-2B}$,
$\sqrt{g}= e^{2B}/2$, and using (\ref{constraint1}), we obtain
\eq\label{gauge1} if_{z} = \frac{g'\,e^{-\varphi_0}}{2i
\left(1-|\phi|^2\right)^2}
    \left[ \phi^{\star} D_z\phi \right] \,,
\eeq
and its complex conjugate. Using the supersymmetry condition
(\ref{constraint1}) on this equation we find
\eq f_{z} = -\frac{g'\,e^{-\varphi_0}}{2} \frac{1}{(1-|\phi|^2)^2}
[\phi^\star \phi_z + \phi \phi^\star_z] \,. \eeq
which is just the derivative of   (\ref{varconstrI}).

\smallskip

 It is straightforward to check that the Einstein equations for the components
($\mu\nu$), $(z,z)$ and ($\bar z, \bar z$) are automatically
satisfied for covariantly holomorphic scalar fields, constant
dilaton and no warping.
On the other hand, the relation  between the gauge function $f$
and the hyperscalars  (\ref{varconstrI}), together with the
antiholomorphicity condition, implies that the  hyperscalar
equation of motion is also satisfied.

\smallskip

Finally, the ($z,\,\bar z$) component of the Einstein's equations
gives us \eq \frac{1}{2}\,B_{z\bar z} =
    -\frac{D_{z}\phi D_{\bar{z}}\phi^{\star}}{2\left(1-|\phi|^{2}\right)^2}
-\frac14 e^{2B+\varphi_0}\,f^2 \,. \eeq

\ni This equation coincides with the integrability constraint
(\ref{integra}), so once this equation is satisfied, it  ensures
that supersymmetry is preserved.
 The equation provides  a constraint
on the function $B$ that must be satisfied in order to obtain a
solution.
So we have seen that {\it all} the field equations can be obtained
from  the supersymmetry  constraints.

\smallskip

We conclude this section by noticing that the
 supersymmetry constraints  (\ref{varconstrI}) and (\ref{constraint1})
are  analogous to the Landau-Ginzburg equations that describe
linear sigma-model vortices in a supergravity
 setting \cite{dvali,BBS,carlos}.
In our system, we are able to solve exactly the resulting equations of motion.
 The important difference with the usual case
 is in the form of the potential, which, in our case, is required by supersymmetry to
 have a minimum at the origin.
 For this   reason, the solutions that we will find have similarities but
 also significant differences to the usual vortex solutions.

\section{The SuperSwirl}\label{secsusyconf}

\subsection{Determining the solution}

In the last section we obtained the constraints that the geometry
and Killing spinors must satisfy in order to have a supersymmetric
configuration. We find that all the equations of motion are
automatically satisfied, once we impose the supersymmetry
constraints
and we are  left with only one nontrivial equation coming from the
$(z,\bar z)$ component of the Einstein equation:
\eq\label{fefB} \frac{1}{2}\,B_{z\bar z} =
    -\frac{D_{z}\phi D_{\bar{z}}\phi^{\star}}{2\left(1-|\phi|^{2}\right)^2}
-\frac14 e^{2B+\varphi_0}\,f^2\,, \label{Ein4} \eeq

\smallskip
\ni while eq. (\ref{constraint1}) and its complex conjugate are:

\eq\label{phiAz}
\partial_{\bar z}\,\phi = \frac{ig'}{2}\, A_{\bar z}\,\phi\,, \qquad \qquad
\qquad \partial_{z}\,\phi^\star = -\frac{ig'}{2}\,
A_{z}\,\phi^\star \,. \eeq
Defining
\eq
\phi=\psi^{\frac12}\,e^{i \tau}\,,
\eeq the fields  that we have to determine are $\tau$, $B$, and
$\psi$. We start by extracting some information from
(\ref{phiAz}).
  It is simple to show, starting from these formulae,
that, whenever $\psi \neq 0$, the following equations hold
\begin{eqnarray}
\partial_{z\bar{z}} \ln{\psi} &=& \frac{i g'}{2}\,
(\partial_z A_{\bar{z}} - \partial_{\bar z} A_{z} ) \,=\,
\frac{i g'}{2} F_{z \bar{z}}\label{combi1}\\
\partial_{z\bar{z}}\,\tau &=& \frac{ g'}{4}\,
(\partial_z A_{\bar{z}} + \partial_{\bar z} A_{z} ) \label{combi2}
\end{eqnarray}


\ni Notice that the following gauge transformation  leaves
invariant these two equations:
\eqa \label{E:gaugexfmn}
&& A_{z}\,\to A_{z} + \partial_{z} T\,, \\
&& \phi \,\to e^{ig'\,T/2} \phi\,. \eeqa
In terms of the fields $\psi$ and $\tau$ the gauge transformation
is\footnote{This shows that the phase  $\tau$
 can  be absorbed by a gauge transformation, and we can identify the former
 with the latter.  In the following section,
we will see that global constraints fix the structure of the
function $T$, and consequently the phase $\tau$.}
\eqa\label{phixfmn}
&& \psi\,\to \psi\,,\\
&& \tau \,\to \tau + \frac{g'}{2}\,T \,. \eeqa

\smallskip
\ni {\it Determining B}

\smallskip
\ni Remembering  that
\eq\label{eqffp} f = -\frac{g'}{2}
    \frac{e^{-\varphi_0}}{(1-|\phi|^2)}\,\left(=\,-2 i F_{z \bar{z}} \,e^{-2 B}
     \right) \,, \label{C2}
\eeq
\noindent eq. (\ref{fefB}) can be rewritten as

\eq\label{firstB} B_{z \bar{z}}=-\frac{\partial_{z \bar{z}}
\ln{\psi}}{(1-\psi)} - \frac{\partial_{z} \psi \,
\partial_{\bar{z}}\psi}{\psi\,(1-\psi)^2} =
\partial_{z\bar z}[\ln{(\psi^{-1}-1)}] \,.
\eeq

\smallskip

\ni Thus we can integrate this equation, to obtain:

\eq\label{firstuB} e^{B}\,=
\frac{\left(1-\psi\right)}{\psi}\,F^{1/2}(z)
F^{\star 1/2}(\bar{z})\,. \eeq In this way, we have found a direct
relation between $\psi$ and the  metric function $B$, given in
(\ref{firstuB}).
Alternatively, a relation between these two quantities is obtained
comparing eq. (\ref{eqffp}) and eq. (\ref{combi1}). One finds

\eq\label{secB} e^{2B} = \frac{8\,e^{\varphi_0}}
    {g'^2}\,(1-\psi)\,\partial_{z \bar{z}} \ln{\psi}\,.
\eeq

\ni {\it Determining $\psi$}
\smallskip

\ni
 Comparing (\ref{firstuB}) and (\ref{secB})  one obtains the following
differential equation for $\psi$, which must be solved to obtain a
SUSY solution:

\eq\label{defpts} \frac{8\,e^{\varphi_0}}{g'^2}\,
    \partial_{z \bar{z}} \ln{\psi}
\,=\,  \frac{\left(1-\psi \right)}{\psi^{2}}\,F(z) F^{\star}(\bar
z)\,.
 \eeq

\smallskip

\ni If regular enough, the function $F$ can  be re-absorbed into
the two dimensional metric  by a rescaling of the coordinate $z$:
$$
dz \,\to \,F(z)\,dz \,.
$$
For this reason we can set it equal to an arbitrary real
integration
 constant, $\tilde c$, without loss of generality. Thus we can rewrite
(\ref{defpts}) as
\eq\label{defpts1}
    \partial_{z \bar{z}} \ln{\psi}
\,=\, c\, \frac{\left(1-\psi \right)}{\psi^{2}}\,,
 \eeq
where the constant $c$ is  given by
\eq\label{defdc} c=\frac{g'^2 \,e^{-\varphi_0}\,\tilde
c^2}{8}\,. \eeq
The most general supersymmetric solution with the matter content
that we are considering, preserving 4D maximal symmetry,
corresponds to the most general solution to the modified Liouville
equation given  in (\ref{defpts1}).

\smallskip

\ni {\it Determining $\epsilon$}

\smallskip

\ni We can now integrate the Killing spinor equation
(\ref{E:deltapsim}) explicitly. Using (\ref{projectionII},
\ref{constraint1}, \ref{spinconnection}) and (\ref{firstuB}) as
discussed above, this equation gives the solution:
\eq\label{spinorsol}
\epsilon(z,\bar z)= e^{i\tau(z,\bar z)}\, \epsilon_0
\eeq
where $\epsilon_0$ is a constant spinor. This solution indeed
satisfies (\ref{spinors}).

\smallskip
\ni {\it The solution}

\smallskip

An exact solution to equation (\ref{defpts1}) can be obtained by asking that
$\psi$ depends on some real combination of $(z,\bar{z})$, for
example by~\footnote{This choice is equivalent  to asking that the
solution is axially symmetric, as we discuss in the next
Section.}
$$ x \equiv z + \bar{z}\,.$$

In this case, it is simple to show that (\ref{defpts1}) can be reduced to
 a first order differential equation

\eq\label{vardifeq}
\left( \frac{d}{d x} \ln{\psi} \right)^{2} \,=\, c\left(\frac{2 \psi -1}{\psi^2} \right) + \alpha^2\,,
\eeq

\ni
where $\alpha^2$ is a positive real constant~\footnote{The case in which
$\alpha^2$ is negative is discussed in the following.}.
 Eq. (\ref{vardifeq}) can be reassembled in the following way

\eq\label{vardifeq2}
\alpha^{2} \left( \psi+\frac{c}{\alpha^2}  \right)^{2}-
\left( \frac{d}{d x} \psi \right)^{2} \,=\,
\left( c + \frac{c^2}{\alpha^2} \right)
\eeq

\ni At this point, it is easy to show that the general solution
for the equation (\ref{vardifeq2}) is given by

\eq\label{soluzione} \psi=\frac{1}{e^{\alpha x}
  }
\,\left[M + N \,
 e^{\alpha x}
         + P\,e^{2 \alpha x}   \right] \,,
\eeq

\noindent where the real numbers   $M$, $N$, $P$
 are integration
constants~\footnote{One can also consider a physically distinct
 solution in which
 $\alpha$, $M$,  $P$ are complex numbers, in a way that  ensures
 that $\psi$ is real. For example, in expression (\ref{soluzione}) one can
 take $\alpha=i \tilde{\alpha}$,
$M=A+iB$,  $P=A-iB$, with  $\tilde{\alpha}$, $A$, $B$ and  $N$
real numbers. This corresponds to the case $\alpha^2$ negative mentioned earlier.
 The global properties of the  resulting
 solution are identical to the one we are going to analyze, and for this
 reason we do not consider this solution in the following.}
 that satisfy  the condition
\begin{eqnarray}\label{relation}
N = - \frac{c}{\alpha^2 } = \frac{1}{2}\left(1- \sqrt{1+ 16 M P}
\right)\,.
\end{eqnarray}
Since $\psi$ is real and positive, this implies that $M,\,P \ge
0$.

\subsection{Properties of the SuperSwirl}

\subsubsection{Axial symmetry}

The  general supersymmetric solution  above,
eq.~(\ref{soluzione}), can be seen to constitute the most  general
axially  symmetry solution that preserves supersymmetry, and
maximal space-time symmetry in 4D.
  This becomes  evident after
performing  the following change of coordinates
$$
e^{2z}\,=\,r e^{i \theta}\,\qquad,\qquad
            e^{2\bar{z}}\,=\,r e^{-i \theta}\,.
$$
that allows one to identify the variable $x$ of the previous
section with $\ln{r}$. The general   solution depending on the
variable $x$, determined in the previous section, in these
coordinates depends
 only on  the radial coordinate
$r$, and, consequently,  it is axially symmetric.

\smallskip
 More explicitly,  a simple calculation shows that in
 terms of these new coordinates,
 the solution~\footnote{Here we are using the gauge freedom to choose
$\tau = g'\,T/2$ (see eq.~(\ref{phixfmn})).} now reads (a $'$ means
derivative along $r$)

\eqa d s_{6}^{2} &=& \eta_{\mu \nu}\, d x^{\mu} dx^{\nu} + e^{2
B(r)} \left( d r^{2}+  r^{2} d \theta^{2}
\right)\,,\label{metrica}\\
 \phi &=& \psi^{\frac12}e^{i\,g'\,T / 2}\,,\label{hyper}\\
\varphi &=& \varphi_0\,,\label{forimfa}\\
 F_{r\theta} &=& -\frac{g'\,e^{-\varphi_0}\,\tilde c^2}{8}
            \frac{(1-\psi)}{r\,\psi^2}\,,\label{flujo} \\
  A_\theta &=&   -\frac{r}{g'}\frac{\psi'}{\psi}+\partial_{\theta}
 T\,.\label{axialgp}
\eeqa

\smallskip

\noindent with the definitions and constraints (notice that we
have redefined the function $B$ as the conformal factor for the
internal metric in polar coordinates;  $e^{2B}dz d\bar z
\rightarrow e^{2B} (dr^2 + r^2 d\theta^2)$):

\eqa
 e^{2B} &=& \frac{\tilde c^2}{4}\, \frac{(1-\psi)^2}{r^2\,\psi^2}\,,  \\
  \psi &=& \frac{1}{r^\alpha}\,\left(M -
    \frac{c}{\alpha^2 }\, r^\alpha + P\, r^{2\alpha} \right)\,, \label{psir}
         \\ \nn \\
c &=& \frac{g'^2 \,e^{-\varphi_0}\,\tilde c^2}{8}=
\,\frac{\alpha^2}{2}\left(
                             \sqrt{1+16MP} -1  \right).
\eeqa

\bigskip

\subsubsection{Singularity structure }

The singularity structure can be read from the metric function
$e^{2B}$ given in formula (\ref{fndB}). When the hyperscalars
are turned on, the solution has
unavoidable (see Appendix), timelike singularities (the scalar
invariants diverge) at the points at which this function vanishes,
or diverges.
 This occurs at the positive zeros of the function $1-\psi=0 $,
where the conformal factor $e^{2B}$ vanishes.  These are located
at
\begin{eqnarray}
r_{\pm}^{\alpha}\,&=&\,\frac{1}{2P}\,\sqrt{\frac{1+\sqrt{1+16
MP}}{2}}\,\left(
\sqrt{\frac{1+\sqrt{1+16 MP}}{2}}\pm1 \right) \\
&=&\,\frac{1}{2P}\,\sqrt{1+\frac{c}{\alpha^2}} \,\left(
\sqrt{1+\frac{c}{\alpha^2}}\pm1 \right) \, .
\end{eqnarray}

\FIGURE[h]{{\epsfig{file=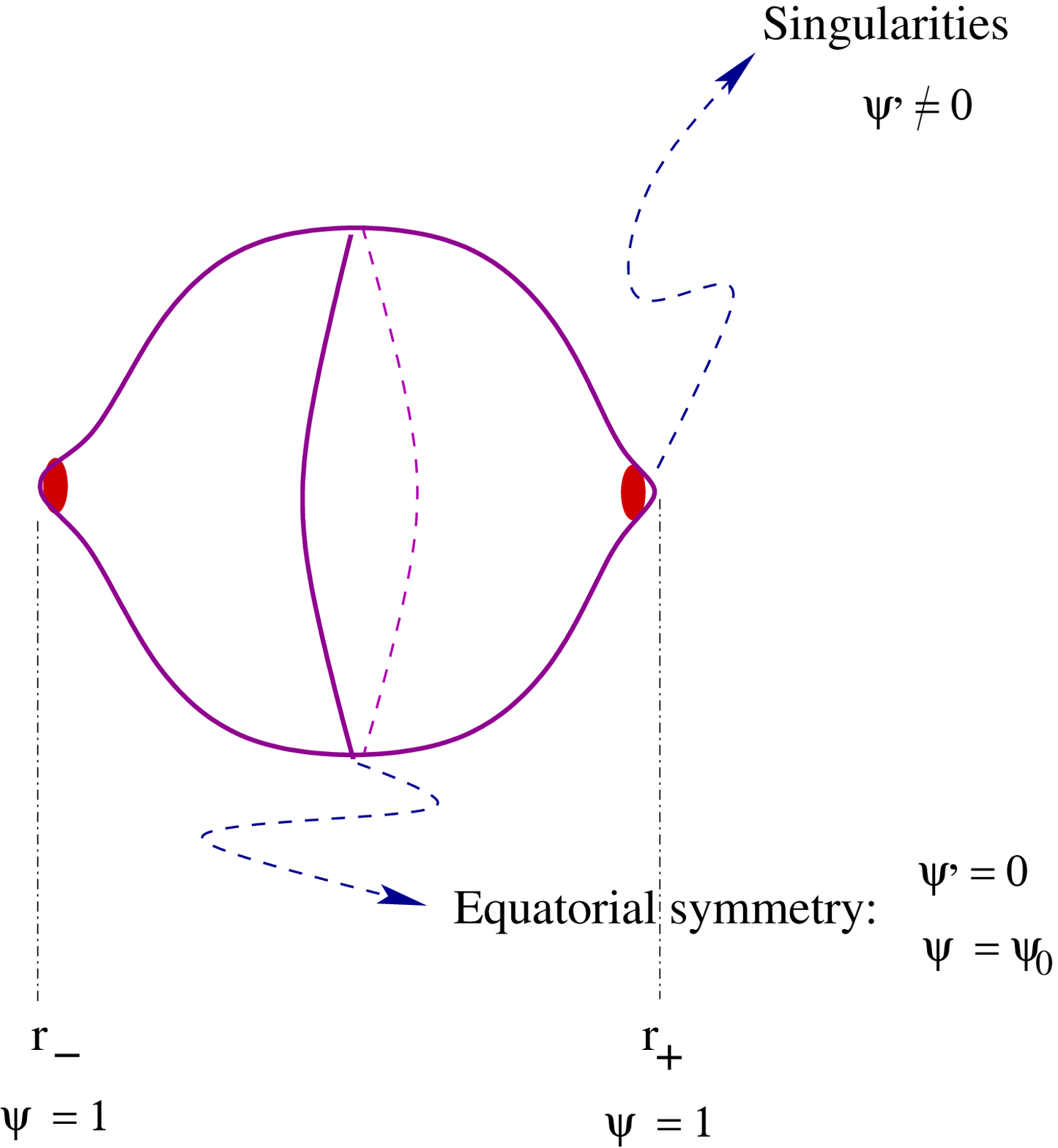, width=.5\textwidth}}%
\caption{Pictorial representation of the internal two dimensional
geometry of the {\it superswirl}. The two points at the ends of
the space are singular. These  are located at the positions where
$\psi=1$. The first derivative -- or speed -- of the scalar field,
does not vanish
 $\psi_\pm'\ne 0$ at these points.
 There is an equatorial symmetry at the position $r_0$, where the
scalar field's first derivative vanishes $\psi_0'=0$. This picture
can also be understood from the point of view of the potential, in
an interesting way, see fig.~(\ref{potential}).\label{geometry}}}

\ni The presence of these singularities is perhaps not surprising,
since the 6D potential and target-space metric, blow up at these
positions. The physical space-time lies in the coordinate range
$r_{-} \le r \le r_{+}$.
 Let us now show how the singularities arise in this spacetime. Consider for example
 the limit $r\to r_{-}$. The relevant part of the metric is
 \eq\label{aume}
 d s_{2}^{2}\,=\,e^{2 B(r)}\,\left( d r^2 + r^{2} \,d \theta^2 \right)\,,
 \eeq

\ni with $e^{2B}$ given in eq. (\ref{fndB}).
Performing  the coordinate transformation

\eq r^{\alpha}\,=\,\sqrt{\rho}\,\sqrt{\frac{4}{\tilde c}
    \,\frac{\alpha\,r_{-}^{\alpha}}{\left( r_{+}^{\alpha}-
r_{-}^{\alpha}\right)}}\,+\,r_{-}^{\alpha}\,, \eeq

\ni brings the metric (\ref{aume}), for $\rho \to 0$ (that is, $r
\to r_{-}$), to the form

\eq d s_{2}^{2}\,\sim\,d\rho^{2} + \gamma\,\rho \,d \theta^2\,,
\eeq

\ni with $ \gamma =4\,\tilde c\,\alpha\,r_-^{\alpha}(r_-^\alpha -
r_+^\alpha)$.
 This implies that near
$r_{-}$ the metric does {\it not} have a conical singularity, but
a more
 serious one.

Notice that the space still closes off on approaching the
singularity, in the sense that a circumference that surrounds the
singularity
 reduces its radius when approaching it.
The  same is true for the limit $r \to r_{+}$.  Moreover, a simple
calculation shows that the internal manifold has a finite volume.


The singularities constitute sources for the hyperscalars. Indeed,
the field $\psi$ and its first derivative do not vanish on
approaching the end of the space at the singularities
 $r_{+}$, $r_{-}$: consequently, a source  producing these fields, with the
right boundary conditions, should be  located at the position of
the singularities.

\subsubsection{Global constraints}

One can see from the expressions for the solution in
eq.~(\ref{axialgp}) that the gauge field strength vanishes at the
position of the singularities. This indicates that the sources are
not coupled to the magnetic field.  We therefore ask that the
gauge potential also be vanishing at the
 singularities.  The gauge field is given by
\eq A_{\theta}(r)\,=\,-\frac{r}{g'}\frac{\psi'}{\psi}+a \,, \eeq
where  $a$ is an integration constant corresponding to the gauge
freedom, that is,  choosing the function $T$ in (\ref{axialgp}) as
 \eq T\,=\,a\, \theta + 2b/g'\,,\label{defofT} \eeq
with $a$ and $b$ real numbers.

In order to have a gauge field  vanishing
 at both $r=r_{\pm}$, it must be defined locally over two overlapping patches,
 with two different integration constants:
\eqa\label{as} a^+\,=\,\frac{r_{+}}{g'}\,\psi'(r_{+} )\,;\qquad
\quad a^-\,=\,\frac{r_{-}}{g'}\,\psi'(r_{-} )\,.\nn \\
\eeqa
Since the hyperscalars are charged under the gauge field
(\ref{E:gaugexfmn}), they must also be locally defined:
$\phi^{\pm} = \psi^{1/2}(r) e^{i \tau^{\pm}}$, with $\tau^{\pm} =
g'a^{\pm}\, \theta/2 + b$ .
 Since $\phi$ must be single-valued over the period $\theta = (0,
2\pi)$, there is a constraint on the integration constants:
\eq \label{singlevalued} \frac{g'}{2} \, a^{\pm} = n^{\pm} \,\,;
\qquad \qquad  n^{\pm} \in {\mathbb Z} \,. \eeq
Inserting  the values for $a^\pm$ in (\ref{as}) into this
expression, we find a topological condition on the parameters of
the solution:
\eq \frac{\alpha}{2}\,\sqrt{1+\frac{c}{\alpha^{2}}}\,=\,n^+ = -
n^-\,.\label{topocondi} \eeq
We thus see that the total winding inside the internal space
vanishes
as it should. %
Moreover, we also require that $A^{\pm}(r)$ (and
$\phi^{\pm}(r,\theta)$) are related in the overlap by a
single-valued gauge transformation: \eqa
A^+ &=& A^- + \partial_{\theta} \Lambda \\
\phi^+ &=& \phi^- \, e^{i \frac{g'}{2}\Lambda} \,, \eeqa
which leads to a Dirac quantization condition.  Given
(\ref{singlevalued}), we see that the conditions above are  indeed
satisfied:
\eq\label{diffintco}
 a^- - a^+\,=\, \frac{ 2 m}{g'} \,,
\eeq
where $m \in {\mathbb Z}$.  From here we immediately find that $m
= 2n^+$.

The previous discussion indicates that the global constraints on
the gauge fields generate a  winding of the hyperscalars around
the singularities. These fields, on each side of the equator are
given by \eq \phi^{\pm} = \psi(r)^{1/2} \, e^{i (n^{\pm}\theta +
b)} \,, \eeq with $ n^{\pm}$ integer numbers, and they smoothly
join at the equator where $\psi'$ vanishes.

\smallskip
\smallskip

 We conclude   this subsection returning to the issue of
supersymmetry for our solution.
  Plugging the superswirl
solution with the global constraints we have just discussed, into
(\ref{spinorsol}),
 we obtain the explicit solution for the Killing spinor, which is
given by:
\eq\label{solkilspi} \epsilon
\,=\,\epsilon_{0}\,e^{i(n^{\pm}\theta+b)}\,, \eeq

\noindent where $\epsilon_{0}$ is, again, a constant spinor.
 From this expression, we explicitly show that the Killing spinor
for our configuration
 is {\it single valued},
since after an interval of $2 \pi$ (the period of the $\theta$
coordinate) the spinor (\ref{solkilspi}) returns to itself.

\subsubsection{The rugby ball limit}

We now show that in the limit when the hyperscalars go to zero in
a proper way,
 we recover the rugby ball solution \cite{ss,ABPQ}. Such a limit,  $\psi\to0$, is
obtained by properly sending $M$, $c$ and $P$ to zero. The
function $e^{2B}$ can be rewritten as
\eq\label{fndB} e^{2B}=\frac{2\, c\, e^{\varphi_0}}{M P \,g'^{2}}
\, \frac{1}{r^2} \, \frac{\left(r^{\alpha}- M+\frac{c}{\alpha^2}
r^{\alpha} -P r^{2\alpha}\right)^2}{ \left[ \left( \frac{M}{P}
\right)^{\frac12} -\frac{c}{\sqrt{M P}\,\alpha^2}r^{\alpha}+
\left( \frac{P}{M} \right)^{\frac12} r^{2\alpha}\right]^2} \eeq

\ni From eq.(\ref{relation}) we learn that, when  $M$ and $P \to
0$, \eq c\to0 \,\qquad\,,\,\qquad\,\frac{c}{\sqrt{MP}} \to
0\,\qquad\,,\,\qquad\,\frac{c}{M P} \to 4\alpha^{2}\,.\eeq

\ni So eq. (\ref{fndB}) becomes, if  $M$ and $P \to 0$ at the same
rate,

\eq\label{sndB} e^{2B}=\frac{8\, e^{\varphi_0}\, \alpha^2}{g'^{2}}
\, \frac{1}{r^2} \, \frac{1}{ \left[
\left(\frac{r}{r_0}\right)^{\alpha} +
\left(\frac{r}{r_0}\right)^{-\alpha} \right]^2}\,, \eeq

\ni with \eq\label{defr0}
r_{0}=\left(\frac{M}{P}\right)^{\frac{1}{2\alpha}}\,, \eeq

\ni and this is nothing but the rugby ball in non-standard
coordinates. In order to see this explicitly, one makes the change
of coordinates: \eq \left(\frac{r}{r_0}\right)^\alpha =
\tan{\frac{\chi}{2}} \,. \eeq
In these coordinates the two dimensional metric becomes: \eq ds^2
= a_0^2 (d\chi^2 + \alpha^2 \sin^2{\chi} \,d\theta^2) \eeq
where $a_0^2= 2 e^{\varphi_0}/ g'^{2}$ is the radius of the
2-sphere and $\alpha$ is related to the deficit angle. One can
similarly check that the gauge field also acquires the right
monopole limit~\footnote{In the rugby-ball limit in which
 the 
hyperscalars go to zero, supersymmetry is  generally broken by
the presence of the deficit angle,
due to the global constraints discussed above.
 See \cite{ABPQ} for details.}.

\subsubsection{Equatorial symmetry}

 Although singular, our solution enjoys an equatorial
symmetry similar  to the rugby ball one: the solution has a reflection symmetry
 on a  hypersurface that we can call the {\it equator}.
 For the rugby ball  in
the coordinates of eq. (\ref{sndB}), it is simple to see that the
equatorial symmetry is translated to the symmetry $r \to
\frac{r_{0}^{2}}{r}$. The point  $r_0$ is the position of the
equator, and is a fixed point for the reflection symmetry. In our
case {\it exactly} the same is true.
 It is indeed simple to show that both the scalar
 (\ref{psir})  and the metric
(\ref{aume})  are invariant under the operation
\eq r\,\to\,\frac{r_{0}^2}{r}\,,\eeq

\ni with the same $r_{0}$ given in (\ref{defr0}). We illustrate
the global structure of the solution in Fig. (\ref{geometry}).

\subsubsection{Energy}\label{energia}

We can now compute explicitly the energy per unit four dimensional
volume of the superswirl. As expected, the energy turns out to
diverge, due to the contributions from the boundaries. Indeed, the
energy can be computed from (see e.g. \cite{dvali})
\eqa {\mathcal E}& =
&\int{dr\,d\theta}\,\sqrt{g}\,\left[\frac{1}{4}R +
     \frac{D_m\phi\, D^m\phi^\star}{(1-|\phi|^2)^2} + \frac{1}{4}
    e^{\varphi_0}F_{mn}F^{mn} + \frac{1}{8}\frac{g'^2\,e^{-\varphi_0}}
                {(1-|\phi|^2)^2}  \right] \nn \\\nn\\
  && 
    +\, \frac{1}{2}\left(\int{d\theta\,\sqrt{h}\,{\mathcal K}}|_{r=r_+}
        - \int{d\theta\,\sqrt{h}\,{\mathcal K}}|_{r=r_-} \right) \,,
\eeqa
where ${\mathcal K}$ is the extrinsic curvature of the surfaces
$r=constant$, whose metric is $h$. In our case these surfaces are
the ``boundaries'' at $r_\pm$. For our solution
(\ref{metrica}-\ref{flujo}) this energy can be expressed in a
Bogomol'nyi type form as follows:
\eqa\label{energy1} {\mathcal E}& =&\frac{1}{2} \int{dr\,d\theta
\,\frac{1}{r} \left[
    \frac{\left|\,r\,D_r\phi+ i\, D_\theta\phi\, \right|^2}{(1-|\phi|^2)^2}
  + e^{\varphi_0}\left(f+ \frac{g'\,e^{\varphi_0}}{2\,(1-|\phi|^2)}\right)^2
     \right] } \nn \\ \nn \\  && 
+  \frac{1}{2}\left(\int{d\theta\, r\,B' }|_{r_+} -
              \int{d\theta\, r\,B' }|_{r_-} \right)\,.
\eeqa Here we have used the ($z,\, \bar z$) component of
Einstein's equations (or the gravitino integrability constraint)
to express $R$ in terms of the matter fields. From this expression
is clear that the supersymmetry constraints
 (\ref{varconstrI})
and (\ref{constraint1}) in terms of the $(r,\, \theta)$
coordinates,
 imply the vanishing of the first two terms of the energy.
 Thus the energy is given entirely by the last two terms. These are
given by
\eq\label{energy2} {\mathcal E}= -\pi\, \left(\frac{r\,
\psi'}{1-\psi} +
     \frac{r\,\psi'}{\psi}\right)\Bigg |_{r_+}
   +\pi\, \left(\frac{r\, \psi'}{1-\psi} +
     \frac{r\,\psi'}{\psi}\right)\Bigg |_{r_-} \,.
\eeq Here we have used the explicit expression for the derivative
of $B$ in terms of $\psi$. It is simple to see that,  at the
boundaries where the curvature singularities are located, this
quantity diverges, since there $\psi=1$. This signals the
necessity to include explicit source terms for the hyperscalars,
placed at the boundaries. Their presence can contribute with new
terms to the calculation of the energy, rendering it finite by
compensating the infinite contributions.

\bigskip

\FIGURE
{{\epsfig{file=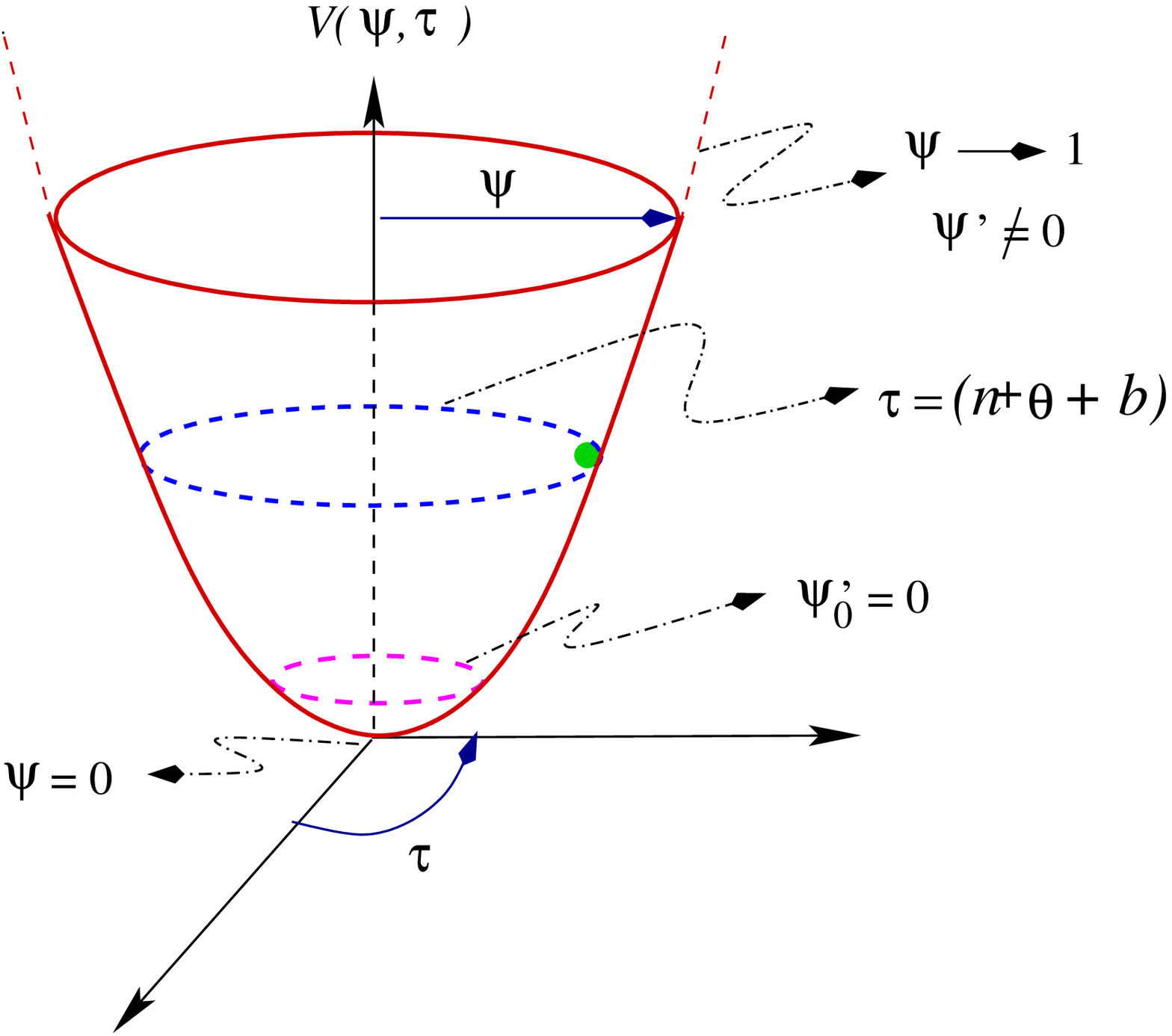, width=.5\textwidth}}%
\caption{The structure of the superswirl solution in terms of the
potential. \label{potential}}}

\section{Discussion}\label{secphysimpl}

We have determined and studied a static,
supersymmetric, codimension-two configuration for a nonlinear
sigma model, in the context of six dimensional gauged
supergravity.
For the matter content considered (whose bosonic part is a $U(1)$ gauge field, and a complex scalar field), it is the most general supersymmetric solution consistent with 4D maximal symmetry and axial symmetry in the internal space.
The solution can be regarded as a deformation of the
classical spherical compactification of Salam-Sezgin, due to a non
trivial profile for the hyperscalars in the internal manifold.
 Although the internal manifold is non-compact, since the presence
of hyperscalars produces singularities at the poles~\footnote{By
{\it poles} we mean the points in which the $(\theta \theta)$
component of the metric vanishes.} of the geometry, it has a
finite volume.  The configuration is everywhere locally
supersymmetric, except at the position of the singularities,
$r_{\pm}$.

The presence of the singularities is an  essential ingredient that
allows the solution to exist: the singularities behave, indeed, as
{\it sources} for  the hyperscalar fields.  Without these
singularities, the complex hyperscalar field, by
 continuity, would need to vanish at the position of the poles  of
the compact manifold.  This is because at the poles, the angular
coordinate, along which the hyperscalar winds, is ill-defined. The
presence of sources where the singularities are located, instead,
allows for more general boundary conditions for the hyperscalars
at the poles, and permits supersymmetry to be preserved away from
the sources.

The exact  supersymmetric solution that we have found has some
similarities with the Landau-Ginzburg (LG) vortices studied in
\cite{BBS,carlos} in 3 dimensions, as well as with the recent
D-strings in 4 dimensions studied in  \cite{dvali}. Indeed, the
conditions required to preserve some fraction of the
supersymmetry,
have the very same structure (see eqs.~(\ref{varconstrI}) and
(\ref{constraint1})).
However, there are important differences between the LG vortices,
the D-strings and our configuration. In the former
cases, the sigma model considered is a linear one, and corresponds
to the
 (non)-abelian Higgs model with a Mexican hat potential. This allows one to
find smooth string solutions, with a well defined size for the
core of the string, which depends on the inverse of the  vacuum
expectation value for the Higgs field. The scalar that generates
the vortex vanishes by continuity at the origin, where the maximum
of the potential is located, and asymptotically, it
 approaches the minimum  of the potential outside the core of the
vortex. The field  has a winding around the symmetry axis,
parameterized by an integer number $n$.  This measures the tension
of the string as seen from infinity, and represents a topological
charge that ensures the stability of the system. Moreover, the
 tension of these strings  is finite, as the boundary terms provide
a finite contribution to it. Indeed, cosmic strings do not have
any sources for the scalar fields,
 and thus, they are completely smooth and
stable.

Our solution shares the property of the winding of the vortex.
Here, the winding of the hyperscalars around the symmetry axis is
parameterized by two integers $n^\pm$, which define the phase of
the field, $\tau$. The integers $n^\pm$ are related to the Dirac
quantization condition that the gauge potentials must satisfy, and
they are equal and opposite.
Thus, the total winding number vanishes, indicating a cancellation
of the total charge inside the 2D internal space.  This is also
analogous to what happens in systems with vortex-anti-vortex
pairs, in compact spaces.

Beyond the winding, however, the configuration constructed in this
paper, has a somewhat different physical interpretation to
conventional vortices. The underlying potential has a minimum at
the origin, and has a paraboloid-like shape, diverging when
$|\phi|=\psi^{\frac12}=1$ (see figure (\ref{potential})). The
hyperscalar configuration that we determined, consequently does
not have a core at the origin, but it is instead generated by the
sources at the ends of the space, $r_\pm$, corresponding to the
circle at $\psi = 1$ where the potential diverges. It extends from
$\psi=1$ to a value $\psi_0 \equiv \psi(r_0) < 1$, which is
characterized by the fact that $\psi'(r_0)=0$ ($\psi^\prime $
changes sign at $r_0$). In some sense, at that point the
hyperscalar turns back and returns up the potential towards the
source.  The point $r_0$ corresponds, not surprisingly, to the
position of the equator of the two dimensional internal manifold
$M_2$. Indeed,  we have shown that our system, with the
hyperscalars included, is ${\mathbb Z}_2$ symmetric at the
equator.

Finally, another important difference in our solution is the fact
that the energy (per unit volume) is infinite, since it is
proportional to the boundary terms computed at the singular
points.  This again indicates the fact that our system, contrary
to the usual vortices, should have boundary source terms  that
cover the singularities. These should regularise the latter,
rendering the total energy finite.
  For these reasons, our configuration,
although similar in many aspects to the usual supersymmetric
vortices/strings, possesses important differences.  Given its
novelty, we name it: {\it the SuperSwirl}.

\smallskip
This new solution constitutes a new class of supersymmetric vacua
for 6D chiral gauged supergravity, with possible implications for
a deeper understanding of the theory itself, in particular its
origin from higher dimensional supergravities or string theories.
A
 string realization of the Nishino-Sezgin (NS) gauged supergravity \cite{NS}
has been found in \cite{CGP}. Unfortunately,
   the hypermultiplet sector  of the theory was not
 considered in their analysis.
 An alternative route for obtaining NS gauged
 supergravity is being developed in \cite{susha}.  In any case, it
would be nice to understand  whether the superswirl has an
interesting
 higher dimensional interpretation in terms of extended objects.

In the context of 6D brane world scenarios, the superswirl
 can provide a natural  setting for a
{\it thick version} of a codimension-two brane world, along
 the lines proposed in
\cite{noi,igjo,ant}. In this case, the singularities  would be
covered by a sort of {\it thick three-brane}.
%
A possibility would be  to place, at  the position of the
singularities or slightly before them, a four-brane on which  the
space ends, characterized by the fact that  one of its spatial
dimensions is compactified on  a circle with small size. The
fields living on the four-brane would be described by  an action
suitably coupled to the bulk fields, that can in principle
 be constructed  along the lines of
 \cite{otto}.  
  In terms of the SLED proposal for the cosmological constant problem, the superswirl is interesting since it provides another class of 4D flat solutions to which the system can evolve, and moreover the only other explicit supersymmetric solution known, apart from that of Salam-Sezgin.
Bulk supersymmetry represents, naturally,  a very
 important property of this model, since
it  contributes to maintaining the bulk stable. In general, we
expect supersymmetry to nevertheless be broken at the position of
the branes, as in the original codimension-two  SLED proposal. It
would be interesting to determine whether our model enables the
construction of
 a brane action that preserves the bulk supersymmetry, for example along the
 lines of  \cite{kallosh}.

\acknowledgments

It is a pleasure to acknowledge enlightening discussions
with  J.~J.~Blanco-Pillado, C.~Burgess,
  S.~de Alwis, O. Dias, R.~Gregory, H.~M.~Lee, D. Mateos,
 F.~Quevedo, S.~Randjbar-Daemi,  F.~Riccioni, J.~Vinet and M.
 Zagermann.
 S.~P.~is funded by    NSERC,
and thanks Cliff Burgess, Rob Myers and the Perimeter Institute for
generous support. G.~T.~is partially supported by the EC $6^{th}$
 Framework Programme MRTN-CT-2004-503369. He also
 acknowledges financial
support from Indiana University, while great part of this work was done.
The work of I.~Z.~was supported by the US
Department of Energy under grant DE-FG02-91-ER-40672.  In addition,
G.~T~and I.~Z.~would like to thank
Perimeter Institute for kind hospitality during the course of
 this work.

\begin{appendix}
\section{Appendix}

Let us  show that it is not possible  to find, in our system,
regular configurations, with hyperscalars turned on,
 without sources for these fields.  We proceed by contradiction.
 Consider
eq. (\ref{firstuB}). This equation, remember, is obtained under
the hypothesis that the $\psi$ field is not zero. The equation
 can be written as ($q$ is a positive constant)

\eq\label{startingsn} \frac{q}{(1-\psi)}=e^{-2B} \partial_{z}
\partial_{\bar{z}}\,\ln{\psi} \eeq

\ni The right hand side of this equation can be written

\eq\label{firstsn} e^{-2B} \partial_{z}
\partial_{\bar{z}}\,\ln{\psi}= \nabla_{M} \nabla^{M} \,\ln{\psi}
\eeq where the indices $M$ run through $z$, $\bar{z}$.

Now, suppose that we  find a solution for our system that
describes a compact, everywhere
regular manifold with no sources for the hyperscalar
fields. This means that you can
integrate both sides of (\ref{startingsn}) over the manifold. The
RHS is zero, since, by (\ref{firstsn}), it is an integral of a
total derivative over a regular,
 boundary-less space. The equation becomes

\eq\label{intesn} \int d z d \bar{z} \sqrt{g} \frac{q}{(1-\psi)}=0
\eeq

Now, recalling that $\psi=|\phi|^2$ cannot be negative, equation
(\ref{firstuB}) shows that the quantity $(1-\psi)$ must be
positive, or at most null, everywhere.

Now, let us return to the integral (\ref{intesn}). Since $
\sqrt{g}=e^{2B}$ is also everywhere positive, the argument of the
integral must be positive. So the integral cannot be equal to
zero, as required.  Therefore, the initial hypothesis that we can
find a compact  manifold everywhere regular leads us to a
contradiction.


\end{appendix}


\begin{thebibliography}{99}
\bibitem{PS1}
For the interested reader, we note here that we can convert from
the conventions of \cite{PS} to those of
 \cite{NS}, by making the replacement: $T^i \rightarrow -T^i$.  Notice
then that the group action on the coset representative $L$: $g \,
L = L \, h$, which in \cite{PS} for infinitessimal transformations
is  $g = 1 + \Lambda^i T^i$, $h = 1 + \Lambda^i \delta^{ix} T^x$,
becomes in \cite{NS} $g = 1 - \Lambda^i T^i$, $h = 1 + \Lambda^i
\delta^{ix} T^x$.
 This is why, in \cite{NS}, the minimal couplings in the
 covariant derivatives of the scalars and fermions appear
with opposite signs, even though we take the generators $T^i$ and
$T^x$ with the same sign.  Finally, notice that the covariant
derivatives for the fermions in \cite{PS} are written $D_M \psi =
\nabla_M \psi + \partial_M \Phi^{\alpha} \omega_{\alpha}^x T^x
\psi + A^i_M C^{ix} T^x \psi$,
 whereas in \cite{NS} they are:  $D_M \psi = \nabla_M \psi +
\partial_M \Phi^{\alpha} \omega_{\alpha}^x T^x \psi - A^i_M C^{ix} T^x \psi$.
\end{thebibliography}
\end{document}